\documentclass[twocolumn,showpacs,prb,amsfonts,amsmath,amssymb,floatfix]{revtex4}
\usepackage{graphicx}
\usepackage{dcolumn}
\usepackage{bm}
\usepackage[usenames]{color}

\begin{document}
\title{GW approximation with LSDA+U method and 
applications to NiO, MnO, and V$_2$O$_3$}
\author{S. Kobayashi$^{1}$}
\author{Y. Nohara$^{1}$}
\author{S. Yamamoto$^{1,2}$}
\author{T. Fujiwara$^{1,2,3,}$}
\email[Corresponding author:\ ]{fujiwara@coral.t.u-tokyo.ac.jp}
\affiliation{$^{1}$ Department of Applied Physics, The University of Tokyo, Tokyo 113-8656, Japan\\
             $^{2}$ Core Research for Evolutional Science and Technology, 
                    Japan Science and Technology Agency (CREST-JST), Japan \\
             $^{3}$ Center for Research and Development of Higher Education, 
                    The University of Tokyo, Tokyo 113-0033, Japan}   
\date{\today}
\begin{abstract}
A GW approximation (GWA) method named U+GWA is proposed, 
where we can start GWA with more localized wave functions obtained 
by the local spin-density approximation (LSDA)+U method. 
Then GWA and U+GWA are applied to MnO, NiO, and V$_2$O$_3$ in antiferromagnetic phase.  
The band gaps and energy spectra show an excellent agreement 
with the experimentally observed results 
and are discussed in details. 
The calculated width of d-bands of V$_2$O$_3$ is much narrower 
than that of the observed one 
which may be a mixture of t$_{2g}^2$ multiplet and single electron  t$_{2g}$ level.
GWA or U+GWA does not work also in the paramagnetic phase of V$_2$O$_3$ 
and the reason for thois is clarified. 
The method of the unique choice of on-site Coulomb interaction is discussed in details. 
The criterion for whether we should adopt GWA or U+GWA is discussed and 
is assessed with the help of the off-diagonal elements of the self-energy. 
\end{abstract}
\maketitle


\section{Introduction}
Electron correlation is important for electronic properties 
in wide variety of materials, where it causes drastic change in physical properties 
in strongly correlated electron systems 
with small change in electron/hole doping or an external field. 
Standard electronic structure calculations are now based 
on the density-functional theory (DFT) and have progressed greatly 
the understanding of electronic structures in condensed matters. 
The polarization function in DFT is derived by the linear response theory of 
homogeneous/inhomogeneous electron gas, 
where the electron-electron correlation is calculated with 
the random-phase approximation (RPA) and effects of static screening are included. 
However, the results could not be free from self-interaction, especially 
in the local (spin-)density approximation (L(S)DA).~\cite{Arai-Fujiwara} 
On the other hand, GW approximation (GWA) does not suffer from 
the problem of the self-interaction 
and can include effects of dynamical screening in the framework of 
many-body perturbation theory.~\cite{Hedin1965}
Moreover, one can estimate the screened Coulomb interaction from the first principles. 
With a recent progress in computational algorithm and computer facilities, 
GWA can be applied  now to realistic materials. 

GWA is the first term approximation of the many-body perturbation series 
and the self-energy is replaced by the lowest order term of 
the perturbation expansion. 
GWA for realistic condensed matters is formulated usually with the LSDA Hamiltonian 
as an unperturbed one without self-consistent calculation 
of the one-body Green's function $G$
and the LSDA exchange-correlation potential is subtracted later 
from the resultant GWA self-energy.  
The GWA self-energy is then given as 
\begin{eqnarray}
 \Sigma=iG_0W, 
\label{SelfEnergy}
\end{eqnarray}
where $G_0$ is the unperturbed Green's function. 
$W$ is the dynamically screened Coulomb interaction 
with the random phase approximation (RPA), 
which is written as 
\begin{eqnarray}
 W = v + v\chi^0 W ,  
\label{dyn-int2}
\end{eqnarray}
where $v$ is the bare Coulomb interaction and 
$\chi^0$ is the lowest order irreducible polarization function $\chi^0 = -i G_0G_0$.
One possible way is that the quasi-particle energy is determined self-consistently 
in the lowest GWA equation with $\chi^0$, 
which is called the ``eigenvalue-only (e-only) self-consistency".~\cite{e-only-SC}
In fact, the e-only self-consistent approach may be good enough  
if LSDA would give reasonably good starting wave functions, 
though it may not be always the case.  
The self-consistent calculation of $G$ in GWA may be another approach, 
but that without the vertex correction leads to unphysical structure of spectra, 
e.g., too wide band width and disappearance of plasmon satellite, 
and violation of the $f$-sum rule, 
though it ensures conservation of particle number and energy and  
then an accurate total energy.~\cite{Barth-Holm1996,Holm-Barth1998}
Other trials of partially self-consistent treatment of GWA, 
named ``the quasiparticle self-consistent" GWA,~\cite{Faleev} 
have been proposed to improve the quasiparticle band structure. 
The essence of these methodologies is in how to obtain localized wave functions 
in transition-metal oxides. 
Another possibility would be the establishment of a methodology 
to start with some unperturbed Hamiltonian which gives localized wave functions 
or correct eigen-energies. 
In the present work, we propose a methodology of GWA, 
starting with wave functions of the LSDA+U method,~\cite{LSDA+U_3rdGen,LSDA+U-LSDA} 
in order to have localized wave functions and 
apply it to several transition-metal oxides, MnO, NiO and V$_2$O$_3$.~\cite{Miyake2006} 
In Sec.~\ref{U+GWA}, the present methodology, named U+GWA, is explained briefly. 
Section \ref{NiO-MnO} is devoted to the application 
to antiferromagnetic insulators (AFI) MnO and NiO. 
Two systems are very typical:  
The electronic structure of MnO can be 
obtained by GWA and, on the contrary, that of NiO by U+GWA. 
The physical reason for this choice is discussed. 
The method of unique choice of the on-site Coulomb interaction $U$ is analyzed in details.  
We show that the starting wave functions by LSDA are satisfactory in MnO and 
are largely improved in NiO by LSDA+U with a proper value of $U$. 
The present U+GWA is then applied to AFI V$_2$O$_3$ in Sec.~\ref{V2O3}.  
Summary and conclusion are given in Sec.~\ref{Last}. 

\section{U+GWA: GW approximation starting from LSDA+U}~\label{U+GWA}

\subsection{U+GWA}

We will present here a theoretical method of GWA with wave functions 
obtained by the LSDA+U method,~\cite{Miyake2006} 
and we apply this method to NiO, MnO, and V$_2$O$_3$ in the proceeding sections. 
Present calculation is based on the linear muffin-tin orbital method with 
the atomic sphere approximation (LMTO-ASA) (Ref.~\onlinecite{LMTO}) and also the LSDA+U method 
with Coulomb interactions of rotational invariance.~\cite{LSDA+U_3rdGen}
The LSDA+U method has shown a reasonable success for many d- and f-electron systems 
in the broad feature of electronic structure.   
Since the LSDA+U method is a kind of the static limit of GWA, 
it could be a good starting approximation of GWA.~\cite{LSDA+U-LSDA} 
The energy gap becomes larger in LSDA+U and the polarization 
becomes smaller because of the energy denominator 
in the RPA polarization function. 
Moreover, wave functions of the LSDA+U method are more localized 
than those by LSDA because of strong on-site Coulomb interaction 
and, therefore, 
can be more preferable for materials with strong electron-electron correlation. 
Once we start from the LSDA+U Hamiltonian, 
the exchange-correlation potential 
and the Hubbard terms of Coulomb interaction should be subtracted 
from the GWA self-energy as 
\begin{eqnarray}
\Delta\Sigma = \Sigma 
                - V^{\rm xc}_{\rm LSDA} - V^{\rm corr}_{\rm LSDA+U},
\label{SE_Correction}
\end{eqnarray}
where $V^{\rm xc}_{\rm LSDA}$ is the exchange-correlation potential 
and $V^{\rm corr}_{\rm LSDA+U}$ is the potential correction derived from 
the Hubbard term in the LSDA+U method.~\cite{LSDA+U_3rdGen} 
This we call LSDA+U+GWA or U+GWA.
In the present formulation, once we set $U=0$ and $J=0$, the last term in 
Eq.~(\ref{SE_Correction}) 
vanishes and  U+GWA is reduced to GWA.
Green's function $G$ is defined as 
\begin{equation}
G(E)=(E-H_0-\Delta \Sigma (E))^{-1}
\end{equation}
and the quasiparticle energy $E_{{\bf k}n}$ 
should be calculated by the self-consistent equation 
(the e-only self-consistency)   
\begin{eqnarray}
 E_{{\bf k}n} = \epsilon_{{\bf k}n} +  {\rm Re}\Delta\Sigma_{{\bf k}n} (E_{{\bf k}n}) ,
\label{GW-band}
\end{eqnarray}
where $\epsilon_{{\bf k}n}$ is the LSDA+U eigen-energy. 
The self-energy correction in Eq.~(\ref{GW-band}) can be written as   
$\Delta\Sigma_{{\bf k}n} (E_{{\bf k}n}) =
 \langle \psi_{{\bf k}n} | \Sigma(E_{{\bf k}n}) -
 V^{\rm xc}_{\rm LSDA} - V^{\rm corr}_{\rm LSDA+U} | \psi_{{\bf k}n} \rangle $. 
We actually carried out energy-only self-consistent calculation 
to satisfy Eq.~(\ref{GW-band}). 
Usually GWA causes an appreciable mixing between the LSDA valence and conduction bands, 
if we include the off-diagonal elements of the self-energy.~\cite{Sakuma2008}
In the present work, the Coulomb $U$ in the LSDA+U calculation 
is introduced in order to obtain localized wave functions. 
The choice of $U$ will be discussed further in Sec.~\ref{Choice_of_U}. 
Our criterion for the value $U$ in U+GWA is that 
the off-diagonal elements of the self-energy become negligibly small, 
which will be discussed in Sec.~\ref{Legitimacy}.
The exchange interaction $J$ is set to be the value of 
the constrained LSDA calculation.~\cite{J-value} 
If the calculation of Green's function were done self-consistently, 
the results would not depend on the choice of values of $U$ and $J$.

\begin{table*}[htb]
\begin{minipage}{\textwidth}
\caption{\label{Gap_moment_W_MnO-NiO} The input Coulomb and exchange interaction 
parameters, 
$U$ (eV) and $J$ (eV), 
static limit of screened Coulomb interaction $W(0)$ (eV), 
direct band gap $E_{\rm G:d}$ (eV), 
indirect band gap $E_{\rm G:id}$ (eV), 
and the spin magnetic moment $M (\mu_\mathrm B)$ for MnO and NiO. 
The calculated direct and indirect band gaps are 
estimated from the calculated quasiparticle energy bands. 
The spherical average values of the bare Coulomb interaction $\langle v \rangle $ are 
23.6~eV in MnO and 27.9~eV in NiO. 
We have found two U+GWA solutions in NiO with $U=1$~eV and 2.0~eV. 
We believe that the solutions connecting to the ones of $U \ge 2.5$~eV continuously 
are correct. See the text in Sub.~\ref{Choice_of_U}.
}
\begin{ruledtabular}
\begin{tabular}[t]{p{1.8cm}|cccccc|cccccc} 
      & \multicolumn{6}{c|}{MnO}& \multicolumn{6}{c}{NiO} \\
      &$U$ & $J$ &$W(0)$&$E_{\rm G:d}$&$E_{\rm G:id}$ &$M$ & $U$ & $J$ & $W(0)$ &$E_{\rm G:d}$&$E_{\rm G:id}$ & $M$ \\ \hline
LSDA  & -   &  -   & 4.50 & 0.80 &0.80 &4.33 &  -  & -    & 1.72  & 0.49 & 0.11 & 1.01 \\
LSDA+U& 1.0 & 0.86 & 4.67 & 0.91 &0.87 &4.35 & 2.5 & 0.95 & 4.14  & 1.79 & 1.43 & 1.52 \\ 
GWA   & -   &  -   & 7.07 & 3.51 &3.05 &4.33 &  -  & -    & 1.51  & 0.38 & 0.21 & 1.13 \\
U+GWA & 1.0 & 0.86 & 7.24 & 3.58 &3.10 &4.35 & 1.0 & 0.95 & 4.43/2.38& 3.52/0.74& 3.05/0.54& 1.34/1.34 \\
      & 2.0 & 0.86 & 8.15 & 3.96 &3.42 &4.45 & 2.0 & 0.95 & 5.54/3.37& 4.10/1.29& 3.47/1.17& 1.44/1.44 \\
      & -   &  -   & -    & -    & -   &-    & 2.5 & 0.95 & 6.03  & 3.97 & 3.46 & 1.46 \\
      & 4.0 & 0.86 & 9.39 & 4.76 &4.18 &4.58 & 4.0 & 0.95 & 7.25  & 4.16 & 3.99 & 1.71 \\
      & 7.5 & 0.86 &10.39 & 4.95 &4.29 &4.71 & 7.5 & 0.95 & 9.43  & 5.48 & 4.78 & 1.71\\
Constrained LSDA\footnotemark 
      & 6.9 &  0.86& -    & -      & -  &- &  8.0& 0.95 & -     &     - &-    & -    \\
exp.  &  -  &   -  & - &\multicolumn{2}{c}{3.6$\sim$3.8\footnotemark} &4.58, 4.79\footnotemark
      &   - & -    & - &\multicolumn{2}{c}{4.0, 4.3\footnotemark}&1.64, 1.77, 1.9\footnotemark\\ 
\end{tabular}
\end{ruledtabular}
\footnotetext[0]{The value of $W(0)$ depends on the orbital components and its averaged one is shown here.}
\footnotetext[1]{Reference \cite{LDA+U-NiO-1}}
\footnotetext[2]{R.N.Iskenderov et al. Sov. Phys. Solid State {\bf 10}, 2031 (1969); 
L.Messick et al. Phys. Rev. B{\bf 6}, 3941 (1972).}
\footnotetext[3]{A.K.Cheetham and D.A.O.Hope, Phys. Rev. B{\bf 27}, 6964 (1983); 
B.E.F.Fender et al. J.Chem.Phys. {\bf 48}, 990 (1968).}
\footnotetext[4]
{S.Hufner et al., Solid State Commun. {\bf 52}, 793 (1984); 
G.A.Sawatzky and J.W.Allen, Phys. Rev. Lett. {\bf 53}, 2339 (1984).}
\footnotetext[5]{H.A.Alperin, J. Phys. Soc. Jpn. Suppl. B{\bf 17}, 12 (1962); 
B.E.F.Fender et al. J.Chem.Phys. {\bf 48}, 990 (1968); 
A.K.Cheetham and D.A.O.Hope, Phys. Rev. B{\bf 27}, 6964 (1983).}
\label{tab:results}
\end{minipage}
\end{table*}

\subsection{Coulomb interactions}
The effective Coulomb and exchange integrals between orbitals 
$\phi_a$ and $\phi_b$, $U(a,b)$ and $J(a,b)$, 
are represented by the Racah parameters $A$, $B$, and $C$  or 
by the Slater integrals $F_0$, $F_2$, and $F_4$.  
The Coulomb and exchange interactions within and between t$_{1g}$ and e$_g$ 
orbitals are as follows in the cubic symmetry:
\begin{eqnarray}
 u_{{\rm t}_{2g}} &=& U(\xi,\xi)= U(\eta,\eta) =U(\zeta,\zeta)\nonumber \\
 &=& u_{{\rm e}_{g}} = U(u,u)=U(v,v)  \nonumber \\
                  &=& A+4B+3C=F_0+4F_2+36F_4,  \\ 
 u^\prime_{{\rm t}_{2g}} &=& U(\xi,\eta)=U(\eta,\zeta)=U(\zeta,\xi) \nonumber \\
                  &=& A-2B+C=F_0-2F_2-4F_4, \\
 u^\prime_{{\rm e}_{g}}  &=&U(u,v)=A-4B+C=F_0-4F_2+6F_4 , \\
 u^{\prime \prime} &=& \frac{1}{6}\sum_{a\in {\rm t}_{2g}}\sum_{b\in {\rm e}_g}U(a,b)=A+C=F_0-14F_4  , \\
 j_{{\rm t}_{2g}}  &=& J(\xi,\eta)=J(\eta,\zeta)=J(\zeta,\xi) \nonumber \\
                   &=& 3B+C =3F_2+20F_4, \\
 j_{{\rm e}_{g}}   &=& J(u,v)=4B+C=4F_2+15F_4, \\
 j^{\prime\prime}  &=& \frac{1}{6}\sum_{a\in {\rm t}_{2g}}\sum_{b\in {\rm e}_g}J(a,b) \nonumber \\
                   &=& 2B+C=2F_2+25F_4 , 
\end{eqnarray}
We have an experimentally observed relationship~\cite{LSDA+U_3rdGen} $F_4/F_2=0.63/9=0.07$   
and the above  expressions are rewritten in terms of the Coulomb and exchange parameters $U$ and $J$ 
with help of the relationships $U=F_0$ and $J=\frac{7}{2}(F_2+9F_4)$ as 
\begin{eqnarray}
u_{{\rm t}_{2g}} &=& u_{{\rm e}_{g}}= U+1.14J, \\
u_{{\rm t}_{2g}}^\prime &=& U-0.40J , \\ 
u_{{\rm e}_{g}}^\prime  &=& U-0.63J , \\ 
j_{{\rm t}_{2g}} &=& 0.77J  , \\
j_{{\rm e}_{g}}  &=& 0.89J  , \\
u^{\prime\prime} &=& U-0.17J , \\ 
j^{\prime\prime} &=& 0.66J . 
\end{eqnarray} 
These $U$ and $J$ may be the input parameters of LSDA+U method or 
the output Coulomb and exchange interactions in U+GWA method.

The energy difference in one-electron spectra 
between an electron affinity level $\Gamma_a$ 
and an electron ionization level $\Gamma_i$ is evaluated as 
\begin{eqnarray}
&& \{E_{\rm T}(\Gamma_a) -E_{\rm T}({\rm G}) \}
-\{E_{\rm T}(G) -E_{\rm T}(\Gamma_i) \} \nonumber \\
&=& {\tilde E}_G + \{E_{\rm C}(\Gamma_a) -E_{\rm C}({\rm G}) \}
-\{E_{\rm C}(G) -E_{\rm C}(\Gamma_i) \}, \nonumber \\
\label{Gap:GW}
\end{eqnarray}
where the total and Coulomb energies of a multiplet $\Gamma$ are written 
as $E_{\rm T}(\Gamma)$ and $E_{\rm C}(\Gamma)$ respectively, 
$E_{\rm T}({\rm G})$ is the total energy of the ground state and
${\tilde E}_G$ is the one-electron energy gap without the Coulomb interactions, 
e.g. a hybridization gap etc. 
In order to evaluate the Coulomb and exchange interactions, we will 
assume, in later sections,  the form 
\begin{eqnarray}
&& \frac{1}{2}\sum_{\{m\} \sigma}[U(m,m^\prime ) n^\sigma_m n^{-\sigma}_{m^\prime}  \nonumber \\
&& \ \ \ \ \ \ \ \ \ \ 
+ \{U(m,m^\prime) - J(m,m^\prime)\} n^\sigma_m n^{\sigma}_{m^\prime} ]  , 
\label{Coulomb}
\end{eqnarray}
where  $n^\sigma_m$ is the electron occupation.

We will analyze the calculated energy spectra by using these formulas 
and show that the ${\tilde E}_G$ vanishes in all these materials 
MnO, NiO and also V$_2$O$_3$.

\section{Electronic structure of NiO and MnO by GWA and U+GWA}~\label{NiO-MnO}

\subsection{Details of calculations}
We use  $4 \times 4 \times 2$ ${\bf k}$-point mesh  
in the Brillouin zone of MnO and NiO. 
The set of the maximum orbital angular momentum of the LMTO basis 
in Mn, Ni, O and empty spheres are chosen to be (ffdp).
In the calculation of the self-energy, the product basis scheme 
is used,~\cite{Ferdi-1,AF-2}
and the maximum total angular momentum of the product bases is set 
to be (ffdp) for the calculation of the correlation part  
$\Sigma^{\rm c}$ of the self-energy.   
This choice reduces the number of the product bases 
used in $\Sigma^{\rm c}$ from 2,376 to 360. 
The e-only self-consistency is achieved with the iterative procedure of 
five or ten times. 
%

\begin{figure}[t] 
\begin{center}
\resizebox{0.48\textwidth}{!}{
\includegraphics[width=8.5cm,clip]{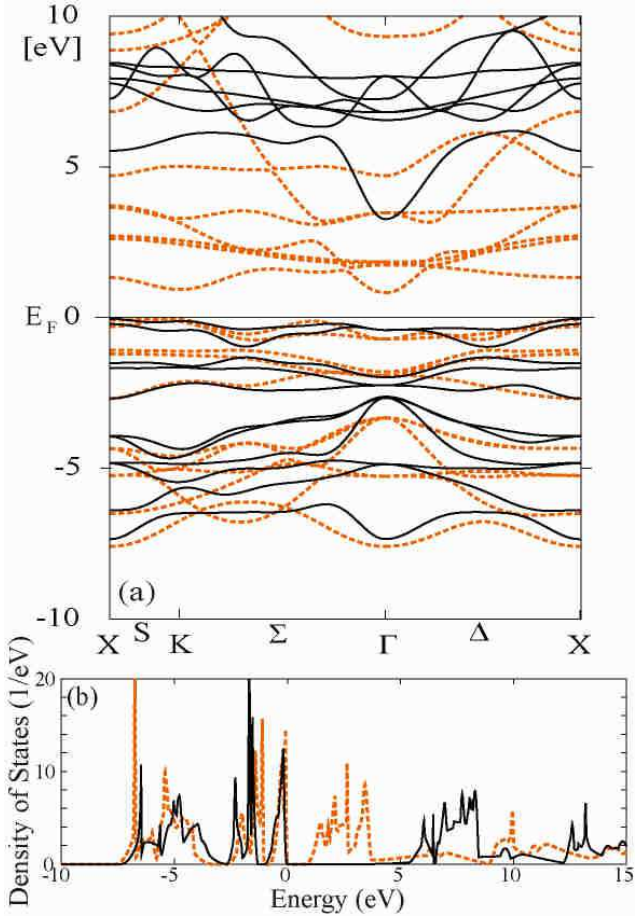}
}
\end{center}
\caption{(Color online) 
(a) Quasiparticle energy bands along symmetry lines 
($\Gamma=(0,0,0)$, X=$\frac{2\pi}{a}(1,0,0)$, K=$\frac{2\pi}{a}(\frac{3}{4},\frac{3}{4},0)$ 
and point $\frac{2\pi}{a}(1,1,0)$ equivalent to X-point) 
and (b) quasiparticle density of states  
of antiferromagnetic MnO.
Solid and broken lines refer to those by GWA and those by LSDA. 
The energy zero-th is fixed at the top of the valence band.    
}
\label{fig:band-MnO}
\end{figure}


\subsection{Electronic structure of antiferromagnetic MnO by GWA}\label{Spec:MnO}

The LSDA band gap of antiferromagnetic MnO is 0.80~eV 
in contrast with the observed one of 3.6$\sim$3.8~eV. 
The observed energy interval between the principal peaks of x-ray photoemission spectroscopy (XPS) 
and bremsstrahlung isochromat spectroscopy (BIS) 
is approximately 9.5~eV and corresponds to 
\begin{eqnarray}
&&\{E_{\rm T}(({\rm t}_{2g}^\uparrow )^3({\rm e}_{g}^\uparrow )^2{\rm t}_{2g}^\downarrow ))
 -E_{\rm T}(({\rm t}_{2g}^\uparrow )^3({\rm e}_{g}^\uparrow )^2)\} \nonumber \\
&&-
\{E_{\rm T}(({\rm t}_{2g}^\uparrow )^3({\rm e}_{g}^\uparrow )^2)
 -E_{\rm T}(({\rm t}_{2g}^\uparrow )^2({\rm e}_{g}^\uparrow )^2)\} \nonumber \\
&&={\tilde E}_{\rm G:MnO}+u_{{\rm t}_{2g}}+2j_{{\rm t}_{2g}}+2j^{\prime\prime} \nonumber \\
&& ={\tilde E}_{\rm G:MnO}+ U+2.57J ,
\label{Gap:MnO}
\end{eqnarray}
where ${\tilde E}_{\rm G: MnO}$ is the hybridization gap in MnO.
The spin polarization in d-orbital is almost complete and 
the calculated numbers of d-electrons of majority and minority spins 
are 4.73 and 0.42, respectively. 
Since the d-d transition for the RPA polarization should be small 
because of almost filled d-band of the majority spin 
and empty d-band of the minority spin. 
Once we adopt GWA for MnO, 
the screening effect on the Coulomb interaction is small and  
the Coulomb interaction cannot be largely screened. 
Therefore, it is expected that 
the energy gap should be open appreciably in MnO with GWA. 

\begin{figure}[t] 
\begin{center}
\resizebox{0.48\textwidth}{!}{
\includegraphics[height=8.5cm,clip]{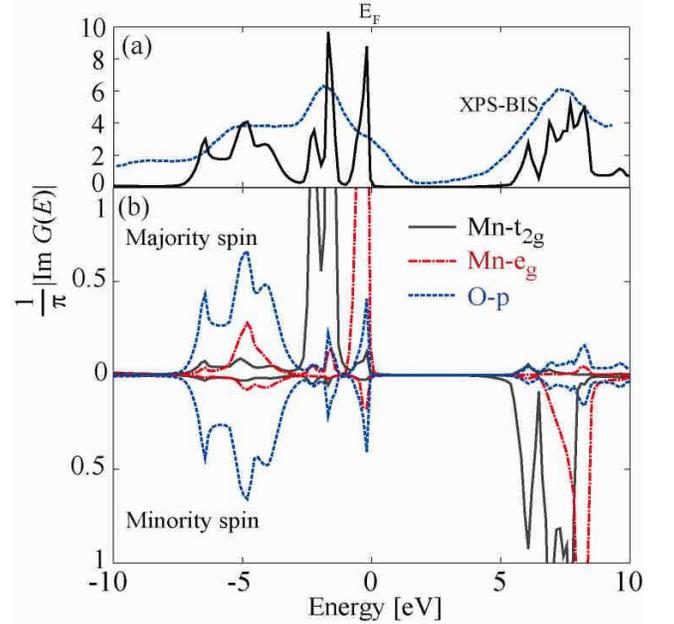}
}
\end{center}
\caption{\label{fig:GreenFn-MnO} (Color online)
Imaginary part of the Green's functions $(1/\pi)|{\rm Im}G(E)|$ 
of antiferromagnetic MnO by GWA.
(a) The total Green's function (solid curve) and 
experimental XPS and BIS spectra (dotted line)~\cite{MnO-exp_XPS_BIS}.  
(b) The partial Green's function per atom.    
The energy zero-th is fixed at the top of the valence band.    
}
\end{figure}

In Fig.~\ref{fig:band-MnO}, the quasiparticle band structure 
by GWA with the e-only self-consistency is depicted 
along the [110] direction from $\Gamma$-point ($\Sigma$- and S-lines) 
and [100] direction ($\Delta$-line). 
It must be noticed that the point $\frac{2\pi}{a}(1,1,0)$ 
along the extended direction of $\Sigma$-line (1,1,0) 
is equivalent to the X-point $\frac{2\pi}{a}(1,0,0)$. 
The corresponding quasiparticle density of states (DOS)  
is shown in Fig.~\ref{fig:band-MnO} (b), 
compared with DOS by LSDA. 
In fact, GWA starting from LSDA gives rise to 
the band gaps to be  $3.05~{\rm eV}$ (indirect gap) and 
$3.51~{\rm eV}$ (direct gap), 
in good agreement with the experimentally observed one 
$E_{\rm G}\simeq 3.6 \sim 3.8~{\rm eV}$. 
The  magnetic moment is evaluated to be $M=4.33~\mu_{\rm B}$ 
comparable to the observed one $M=4.6\sim 4.9~\mu_{\rm B}$.
Therefore, we believe that our expected mechanism for 
less screened Coulomb interaction is the case in MnO. 
The U+GWA is also applied to AFI MnO with the e-only self-consistency 
as summarized in Table \ref{tab:results} in the range of $U=0 \sim 2~{\rm eV}$ 
and $J=0.86~{\rm eV}$ and the spectra and the quasiparticle band structure 
are much the same as those of GWA.  
We will discuss, in Secs.~\ref{Choice_of_U} and \ref{Legitimacy}, 
the choice of $U$ value and the starting wave functions in GWA or U+GWA of MnO. 
Figure \ref{fig:GreenFn-MnO} depicts the imaginary part of the Green's function 
$(1/\pi)|{\rm Im}G(E)|$ by GWA. 
The highest occupied band is the one with strongly hybridization 
between O p and Mn d (e$_g$) states of majority spin 
and the lowest unoccupied one is mainly Mn d (t$_{2g}$) states of 
minority spin. 
This fact tells that 
little d-d transition does contribute to the screening phenomena 
and also that MnO is just in the midway between 
of the charge-transfer type and Mott-Hubbard type insulators.  
The static limit of the screened Coulomb interaction $W(0)$ 
is calculated to be $7.07~{\rm eV}$, 
which means that the screened effect is small. 
We can conclude that GWA can reproduce the energy spectrum 
of AFI MnO satisfactorily, 
since little d-d transition of the polarization 
can participate in the screening phenomena.  

\subsection{Electronic structure of antiferromagnetic NiO by U+GWA}\label{Spec:NiO}

NiO is one of materials where the LSDA wave functions 
may be extended in space.~\cite{Aryasetiawan-NiO-1995} 
The LSDA spectrum shows that the majority spin d-bands are 
almost full and the minority ones are half-full 
and that the band gap $E_{\rm G}$ is 0.11~eV (indirect gap) 
and 0.49~eV (direct gap) 
in contrast with the experimentally observed one of about $4.0\sim 4.3$~eV.

\begin{figure}[t] 
\begin{center}
\resizebox{0.48\textwidth}{!}{
\includegraphics[width=8.5cm,clip]{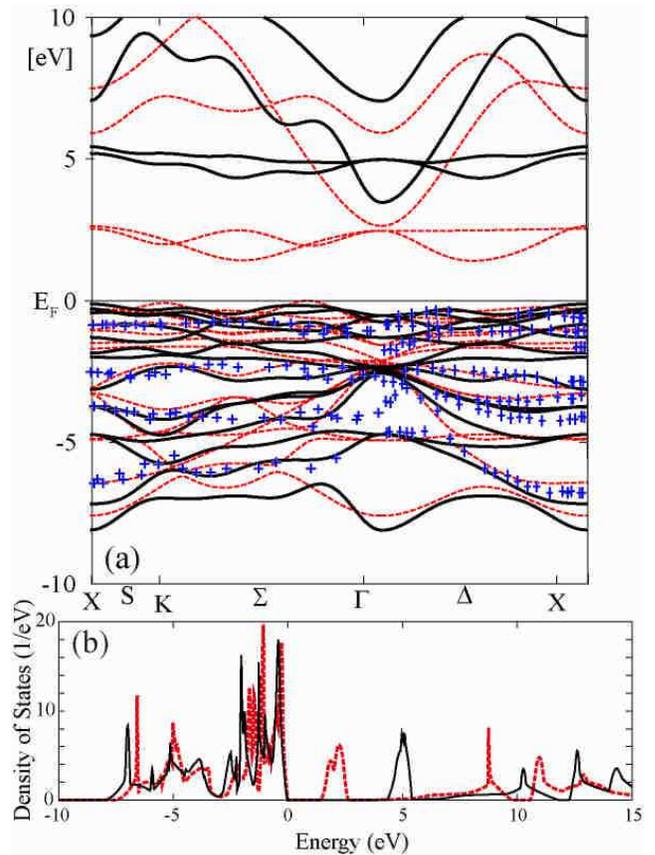}
}
\end{center}
\vspace{-5mm}
\caption{(Color online) 
(a) Quasiparticle energy bands and (b) quasiparticle density of states  
of antiferromagnetic NiO. 
Symmetry points and lines are the same as those in Fig.~\ref{fig:band-MnO}.
Solid and broken lines refer to those by U+GWA and LSDA+U with $U=2.5~{\rm eV}$ 
and $J=0.95~{\rm eV}$, respectively.  
In (a), the cross marks refer to the result by the angle resolved photoemission 
spectra.~\cite{Shen}
The energy zero-th is fixed at the top of the valence band.    
}
\label{fig:band-NiO}
\end{figure}

%
The observed energy interval between the principal peaks of XPS and BIS 
is approximately $6 \sim 6.2$~eV and corresponds to 
\begin{eqnarray}
&&\{E_{\rm T}(({\rm t}_{2g}^\uparrow )^3({\rm t}_{2g}^\downarrow )^3({\rm e}_{g}^\uparrow )^2{\rm e}_{g}^\downarrow ))
 -E_{\rm T}(({\rm t}_{2g}^\uparrow )^3({\rm t}_{2g}^\downarrow )^3({\rm e}_{g}^\uparrow )^2)\} \nonumber \\
&&-
\{E_{\rm T}(({\rm t}_{2g}^\uparrow )^3({\rm t}_{2g}^\downarrow )^3({\rm e}_{g}^\uparrow )^2)
 -E_{\rm T}(({\rm t}_{2g}^\uparrow )^3({\rm t}_{2g}^\downarrow )^3{\rm e}_{g}^\uparrow )\} \nonumber \\
&&={\tilde E}_{\rm G:NiO}+u^\prime_{{\rm e}_{g}}+j_{{\rm e}_{g}}
  ={\tilde E}_{\rm G: NiO}+ U+0.26J ,\nonumber \\
\label{Gap:NiO}
\end{eqnarray}
where ${\tilde E}_{\rm G: NiO}$ is the hybridization gap in NiO.
GWA starting from LSDA gives rise to 
the band gap of $E_{\rm G}\simeq 0.21~{\rm eV}$ 
and the screened Coulomb interaction is estimated as $W(0)=1.51~{\rm eV}$,  
which does not agree with the experimental results. 
This is because the polarization function is too large 
due to small energy denominator of the RPA polarization function.
We would expect a substantial change in wave functions and the polarization function 
led by an increase in the band gap, once we use GWA procedure  
under the condition of opening the gap.~\cite{Aryasetiawan-NiO-1995} 
The LSDA+U method gives better spectrum in NiO.~\cite{LDA+U-NiO-1,LDA+U-NiO-2}

\begin{figure}[t] 
\begin{center}
\resizebox{0.49\textwidth}{!}{
\includegraphics[height=8.5cm,clip]{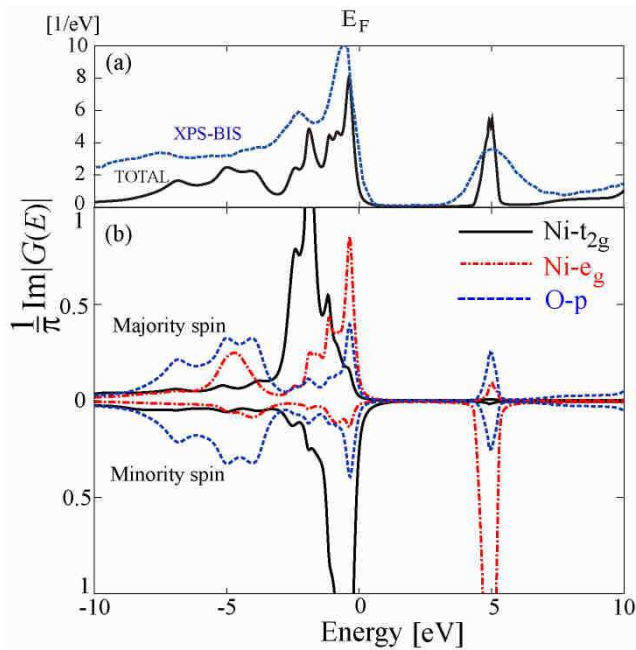}
}
\end{center}
\vspace{-5mm}
\caption{(Color online) 
Imaginary part of the Green's functions $(1/\pi)|{\rm Im}G(E)|$ 
of antiferromagnetic NiO by U+GWA with $U=2.5~{\rm eV}$ 
and $J=0.95~{\rm eV}$. 
(a) The total Green's function (solid line) and 
experimental XPS and BIS spectra (dotted line).~\cite{LDA+U-NiO-2}  
(b) The partial Green's function per atom. 
The energy zero-th is fixed at the top of the valence band.    
}
\label{fig:GreenFn-NiO}
\end{figure}

U+GWA is applied to AFI NiO in the present work 
with the e-only self-consistency and 
a part of the results is shown in Table~\ref{tab:results}, 
compared with those by LSDA, LSDA+U, GWA and experiments. 
In fact, we have used two initial guesses of eigen-energies 
for the e-only self-consistent calculation for $U=1.0$~eV and $U=2.0$~eV; 
one starts from the LSDA+U eigen-energies and 
the other is from the LSDA+U eigen-energies plus some additional shift. 
Then we found different converged solutions for respective initial guesses. 
The cases $U=0$ and $U \ge 2.5$~eV give unique solutions in two starting eigen-values. 
The calculated values of these solutions are summarized in Table \ref{tab:results}. 
We believe that the results connecting continuously to solutions of $U\ge 2.5$~eV should be 
the correct ones and the other ones must be artifact. 
The calculation has been carried out in the range of 
values of $U=0 \sim 8.0$~eV and $J=0.95$~eV (the value by the constrained LDA). 
The calculated peak position (at about 5~eV) of unoccupied bands with $U=2.5~{\rm eV}$ 
is in an excellent agreement with experiments. 
The off-diagonal elements of the self-energy are extremely reduced 
at $U=2.5~{\rm eV}$. 
Then, we adopt the case of $U=2.5~{\rm eV}$ as appropriate choice. 
This choice will be discussed in more details 
in Sec.~\ref{Choice_of_U}.  
The band gaps are  $3.97~{\rm eV}$ (direct) and $3.46~{\rm eV}$ (indirect) 
and the magnetic moment $M=1.46~{\mu_{\rm B}}$ are 
in a good agreement with experimental ones as shown in 
Table~\ref{tab:results}. 
The static limit of the screened Coulomb interaction $W(0)$ is 
6.03~{eV}.
In Fig.~\ref{fig:band-NiO}, the quasiparticle band structure 
by U+GWA with $U=2.5~{\rm eV}$ is depicted 
along the [110] and [100] symmetry lines, 
compared with the angle resolved photoemission spectroscopy data,~\cite{Shen} 
and this shows an excellent agreement between them. 
The d-band width in Fig.~\ref{fig:band-NiO}(a) 
(the energy width of upper occupied eight bands) 
becomes wider  than that of LSDA (but does not change much from 
that of LSDA+U), 
since the occupied O p-bands come upward and the hybridization mixing 
between Ni d and O p becomes larger. 
It should be noted that the Ni 4s-band comes down slightly below 
the flat unoccupied Ni 3d-band. 
The corresponding quasiparticle DOS is shown in Fig.~\ref{fig:band-NiO}(b), 
compared with DOS by LSDA+U. 
Figure \ref{fig:GreenFn-NiO} shows the imaginary part of Green's function 
$(1/\pi)|{\rm Im}G(E)|$ by U+GWA with $U=2.5~{\rm eV}$, 
with the photoemission and inverse photoemission spectra.~\cite{LDA+U-NiO-2}
The most striking feature of  $\frac{1}{\pi}|{\rm Im}G|$ 
is the band width and the intensity distribution in the spectra,  e.g. 
the weight of the spectrum is shifted to the upper region of the occupied bands. 
The weight of O p orbitals in the highest occupied bands increases 
due to the stronger hybridization between O p and Ni d states 
and the lowest unoccupied ones are mainly Ni d states. 
This fact is in a good agreement with observed one and NiO is 
of the charge-transfer type.~\cite{typeNiO}
The screened Coulomb interaction can be calculated in GWA and U+GWA 
as shown in Table~\ref{tab:results}. 
The screening effects are strong in the energy region $0<E < 20~{\rm eV}$ in GWA, 
since all d-electrons participate in the screening phenomena. 
The U+GWA can give much larger $W(0)$ and  
reproduce the energy spectrum of antiferromagnetic NiO very satisfactorily.

\subsection{Unique choice of $U$ in U+GWA in MnO and NiO}\label{Choice_of_U}

We discuss, in this subsection, the $U$ dependence of 
the static screened Coulomb interaction $W(0)$ 
and explain a unique choice of $U$ values in MnO and NiO. 

\begin{figure}[t] 
\begin{center}
\resizebox{0.48\textwidth}{!}{
\includegraphics[height=7.2cm,clip]{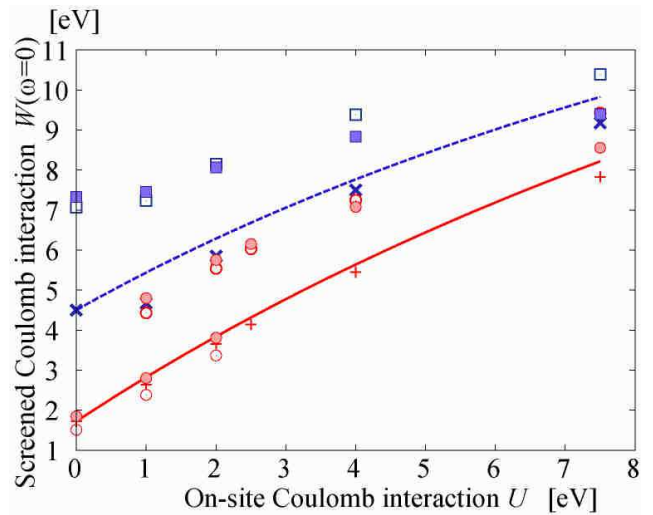}
}
\end{center}
\vspace{-5mm}
\caption{(Color online) 
$U$ dependence of the static screened Coulomb interaction $W_U$. 
$\times$: LSDA+U, open square $\Box$: U+GWA, broken line: Eq.~(\ref{scr-u:LSDA+U})
and closed square: Eq.~(\ref{scr-u:U+GWA}) for MnO. 
$+$: LSDA+U, open circle $\bigcirc$: U+GWA, solid line: Eq.~(\ref{scr-u:LSDA+U}),  
and closed circle: Eq.~(\ref{scr-u:U+GWA}) for NiO. 
See Subsec.~\ref{Spec:NiO} for the explanation on the two solutions of U+GWA 
($\bigcirc$) at $U=1$ and $U=2$ in NiO.
}
\label{fig:W(U)}
\end{figure}

%
The irreducible polarization function in LSDA+U method may be expressed approximately 
in the static limit as 
\begin{eqnarray}
\chi^{0:{\rm LSDA+U}}_U = -\frac{b^2}{\langle E_{\rm G}^{\rm LDA}\rangle+U} ,
\label{Polarization:LSDA+U}
\end{eqnarray}
where $\langle E_{\rm G}^{\rm LDA}\rangle$ is the effective band gap 
of the transition-metal d-bands of the minority spin by LSDA,  and 
$b^2$ is a numerical factor of order 1. 
The static screened Coulomb interaction $W_U$ is then rewritten as 
\begin{eqnarray}
W^{\rm LSDA+U}_U= W^{\rm LSDA}_{U=0} 
\frac{1+\frac{U}{\langle E_{\rm G}^{\rm LDA}\rangle}}
{1+\frac{U}{\langle E_{\rm G}^{\rm LDA}\rangle} \frac{W^{\rm LSDA}_{U=0}}{\langle v \rangle }}  \ ,
\label{scr-u:LSDA+U}
\end{eqnarray}
with the spherical average of matrix elements of the bare Coulomb interaction 
$\langle v \rangle $ and 
$W^{\rm LSDA}_{U=0}=\langle v \rangle /(1-\chi^{0:{\rm LSDA+U}}_{U=0} \langle v \rangle )$.  
We define the effective band gaps $\langle E_{\rm G}^{\rm LDA}\rangle$ to be  
3.7~eV for MnO and 1.4~eV for NiO from the partial density of states 
of the quasiparticle bands. 
We also estimate the $b^2$ values as 
0.665 for MnO and 0.764 for NiO so that 
$W^{\rm LSDA}_{U=0}$ equals to the values of static screened Coulomb interaction by LSDA. 
The $W_U$ values by LSDA+U and U+GWA methods are shown in Fig.~\ref{fig:W(U)}. 
Equation (\ref{scr-u:LSDA+U}) can represent very nicely the values of 
$W_U$ by LSDA+U in both MnO and NiO. 
In the MnO case, the values of LSDA+U method are shifted in the whole $U$ range 
of U+GWA and continuously vary. 
In NiO, on the contrary, there are two solutions at $U=1$~eV and 2~eV, 
and then they vary smoothly. 
To understand the static screening Coulomb interaction of the U+GWA method $W^{\rm U+GWA}_U$, 
we can establish a similar model of the irreducible polarization 
of Eq.~(\ref{Polarization:LSDA+U}) as
\begin{eqnarray}
\chi^{0:{\rm U+GWA}}_U = -\frac{b^2}{\langle {\tilde E}_{\rm G}\rangle+W^{\rm U+GWA}_U} ,
\label{Polarization:U+GWA}
\end{eqnarray}
where $\langle {\tilde E}_{\rm G} \rangle$ is the hybridization gap appearing in Eqs.~(\ref{Gap:GW}), (\ref{Gap:MnO}) 
and (\ref{Gap:NiO}) and $b^2$ is the coefficient already determined 
in Eq.~(\ref{Polarization:LSDA+U}).
Then the static screened Coulomb interaction may be expressed as
\begin{eqnarray}
{\tilde W}^{\rm U+GWA}_U=\frac{\langle v \rangle}{1-\chi^{0:{\rm U+GWA}}_U \langle v \rangle} .
\label{scr-u:U+GWA}
\end{eqnarray}
The right-hand side of Eq.~(\ref{scr-u:U+GWA}) is plotted in Fig.~\ref{fig:W(U)} by substituting 
the calculated static screened constant $W(0)$ by U+GWA in the expression $\chi^{0:{\rm U+GWA}}_U$ 
with  $\langle {\tilde E}_{\rm G} \rangle=0$ 
and shows an excellent agreement between   $W(0)$ by U+GWA and ${\tilde W}^{\rm U+GWA}_U$. 
Of course, though we cannot use the Eq.~(\ref{scr-u:U+GWA}) to determine $W(0)$ or input $U$ value, 
this agreement implies that our treatment is consistent and, furthermore, 
the hybridization gap should be zero in both MnO and NiO.
Though we cannot determine the value of $U$ only in the framework of U+GWA, 
we can do with  help of experimental spectra. 
We replace $U$ in Eqs.~(\ref{Gap:MnO}) and (\ref{Gap:NiO}) with $U_{\rm U+GWA}=W(0)$ 
and we know now the values of the hybridization gap ${\tilde E}_{\rm G}$ equal to zero. 
With the help of the $J$ value of the constrained LSDA calculation $J_{\rm U+GWA}$, 
we have  
$ W(0)+2.57J_{\rm U+GWA} = W(0)+2.21 \simeq 9.5  \ \ i.e.,  \ \ W(0)=7.29$~eV for MnO 
and 
$ W(0)+0.26J_{\rm U+GWA} = W(0)+0.25 \simeq 6.2  \ \ i.e., \ \ W(0)=5.95$~eV for NiO. 
This is perfectly consistent with the results obtained by the choices of 
$U=0$~eV for MnO and $U=2.5$~eV for NiO in U+GWA calculation.

\begin{figure}[t] 
\begin{center}
\resizebox{0.49\textwidth}{!}{
\includegraphics[height=8cm,clip]{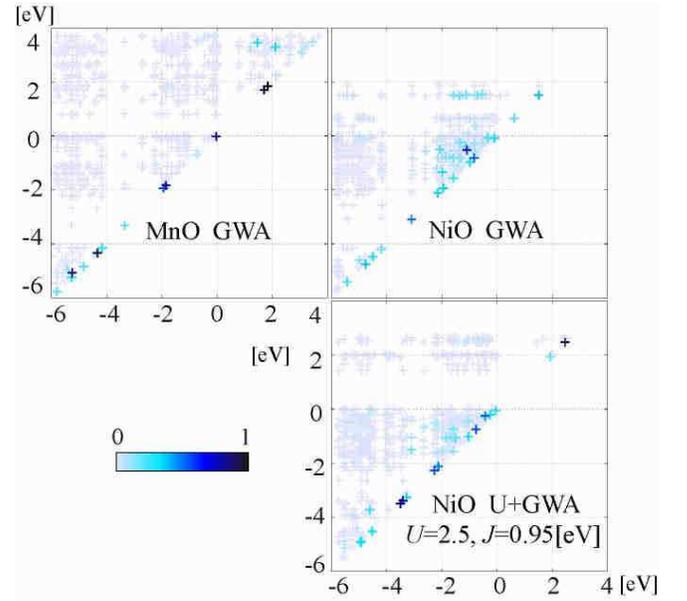}
}
\end{center}
\vspace{-5mm}
\caption{ 
(Color online) The value $(\sqrt{x^2+1}-x)^2$ is shown, whose definition 
should be referred in the text and Eq.~(\ref{Legitimate}). 
The vertical and horizontal axes are the LSDA eigen-energies 
for GWA or LSDA+U eigen-energies (U+GWA) 
$\epsilon_{{\bf k}n}$ and $\epsilon_{{\bf k}n^\prime}$. 
The energy zeroth is set at the Fermi energy. 
The area having larger values in GWA disappears in U+GWA of NiO  
and no area having larger values exists in GWA of MnO. 
}
\label{fig:Legitimacy}
\end{figure}

\subsection{Legitimacy of U+GWA}\label{Legitimacy}

We have shown the calculated results of AFI MnO and NiO 
based on GWA and U+GWA, respectively. 
These two systems correspond to two different situations; \\
(1) The case (MnO) where the d-d transition contributing to the RPA polarization is small. 
The screening effect in GWA of the Coulomb interaction is relatively small and  
the screened Coulomb interaction is not reduced. 
In this case, large improvement can be expected in GWA calculation. \\
(2) The case (NiO) where the d-d transition contributing 
to the RPA polarization is large. 
The screening effect of the Coulomb interaction is large and  
the screened Coulomb interaction is reduced largely in GWA. 
In this case, large improvement could not be expected in GWA calculation 
and we should use U+GWA in order to make the energy gap open.  
Here we show how LSDA+U method improves the wave functions and polarization 
in NiO and 
how LSDA method provides reasonable results in MnO. 
We evaluate the mixing amplitude of the adopted wave functions by the off-diagonal 
element of the self-energy. 
With the off-diagonal matrix element of the self-energy 
$\Delta\Sigma_{{\bf k}nn^\prime}$, 
the two components of bands  $n$ and $n^\prime$ at ${\bf k}$ 
(the LSDA or LSDA+U eigen energies 
$\epsilon_{{\bf k}n}$ and $\epsilon_{{\bf k}n^\prime}$) 
are mixed with a ratio of $(\sqrt{x^2+1}-x)^2:1$,~\cite{Sakuma2008} where 
\begin{eqnarray}
 x = \Big|
\frac{
  \epsilon_{{\bf k}n}       + \Delta\Sigma_{{\bf k}n}(\epsilon_{{\bf k}n}) 
 -\epsilon_{{\bf k}n^\prime}- \Delta\Sigma_{{\bf k}n^\prime}(\epsilon_{{\bf k}n^\prime})  }
 {2 \Delta\Sigma_{{\bf k}nn^\prime} (\{\epsilon_{{\bf k}n}+\epsilon_{{\bf k}n^\prime} \}/2)} 
\Big|     .
\label{Legitimate}
\end{eqnarray}
When this ratio is small ($x$ is much larger than the unity), 
the resultant mixing is negligible. 
The values of this ratio for all pairs of bands 
(\{${\bf k}n$\} and \{${\bf k}n^\prime$\}) 
are shown in Fig.~\ref{fig:Legitimacy}. 
One  can see, in NiO, that the area with larger ratio in GWA disappear in U+GWA 
of $U=2.5~{\rm eV}$. 
On the other hand, in MnO, no area with larger ratio exists in GWA. 
Therefore, in MnO, GWA does not promote a large mixing between 
orbitals contributing to valence and conduction bands. 
On the contrary, 
in NiO, the U+GWA improves the starting wave functions 
and large off-diagonal elements of the self-energy disappear 
between orbitals contributing to valence and conduction bands. 

\section{Electronic structure of V$_2$O$_3$ by U+GWA}\label{V2O3}

\begin{figure}[t] 
\begin{center}
\resizebox{0.4\textwidth}{!}{
\includegraphics[width=7.0cm,clip]{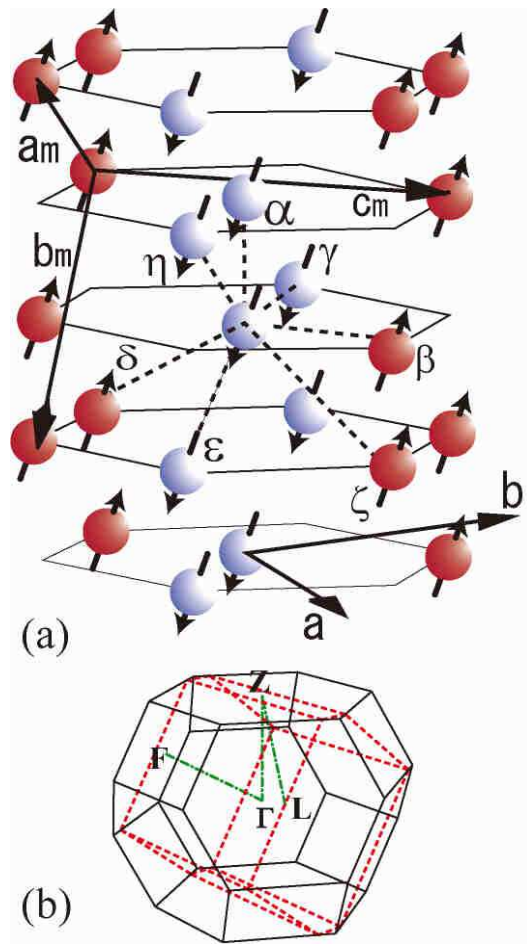}
}
\end{center}
\caption{(Color online) 
(a) Spin structure of low temperature antiferromagnetic insulator phase of V$_2$O$_3$. 
Arrows stands for the spin-directions of V ions. 
The Greek letters $\alpha$, $\beta$, $\gamma$ and $\delta$ denote the first, second, 
third and fourth nearest neighbor V-V pairs, respectively. 
The vectors ${\bf a}$, ${\bf b }$ and their perpendicular vector ${\bf c}$ are the 
primitive vectors of the hexagonal unit cell and, 
the vectors ${\bf a}_m$, ${\bf b }_m$ and ${\bf c }_m$ are those 
of the monoclinic cell of antiferromagnetic phase. 
(b) The Brillouin zone of the paramagnetic (solid lines) and antiferromagnetic 
(broken lines) phases of the corundum structure. The band structure  
in Fig.~\ref{fig:band-V2O3} is 
shown along the symmetry lines indicated by dot-dashed lines. 
The ${\bf k}$-points L, Z and F are on the Brillouin zone of paramagnetic phase 
and F/2 on that of antiferromagnetic phase.}
\label{fig:config-V2O3}
\end{figure}

The antiferromagnetic V$_2$O$_3$ is the other example 
where a strong on-site Coulomb interaction plays a crucial role 
in the electronic structure. 
The crystal structure in the low-temperature AFI phase 
is monoclinic and that in paramagnetic metallic (PM) phase above 150~K is 
corundum structure. 
The spin structure in AFI is depicted in Fig.~\ref{fig:config-V2O3}(a), 
where each V$^{3+}$ ion has one spin-parallel 
($\gamma$ in Fig.~\ref{fig:config-V2O3}(a)) 
and two spin-antiparallel neighbors ($\beta$) 
in the same layer perpendicular to the c-axis 
and one spin-parallel neighbor ($\alpha$) in a different layer.
We can have another view that V$^{3+}$ ions on one layer parallel to the c-axis 
are all ferromagnetically aligned and the inter-layer coupling is 
antiferromagnetic. 
The spin and orbital magnetic moments are observed 
to be $\langle 2S \rangle=1.7~{\mu_B}$ and 
$\langle L \rangle=-0.5~{\mu_B}$ and the total spin is supposed to be $S=1$.~\cite{Paolani}
The observed band gap is $E_{\rm G}= 0.35\sim 0.66~{\rm eV}$.
Under the trigonal symmetry around V ions, 
V d$_{{\rm t}_{2g}}$ level is split into 
non-degenerate a$_{1g}$ and doubly degenerate e$_g^\pi$ levels.  

Mott pointed out that the metal-to-insulator (MI) transition in V$_2$O$_3$ 
at about $150$~K is due to the electron-electron correlations.~\cite{Mott1974} 
The LSDA calculation shows that the a$_{1g}$ states split into bonding 
and anti-bonding and 
that the lowest V d-state is a$_{1g}$ bonding state.~\cite{Mattheiss-1994}  
Then it was proposed that two a$_{1g}$ electrons on a pair of V$^{3+}$ ions 
formed a spin-singlet and 
the remaining e$_g^\pi$ electrons results 
a Mott-Hubbard model of an $S=\frac{1}{2}$ spin on each V$^{3+}$ ion.~\cite{Castellani}
Ezhov {\it et. al.} studied its electronic structure by using the LSDA+U method 
with $U=2.8$~eV and $J=0.93$~eV,   
where the lowest V d-state is e$_g^\pi$ and the a$_{1g}$ is the next lowest. 
They showed that the orbital occupation is predominantly e$_g^\pi$e$_g^\pi$ 
(a spin $S=1$) with a small fraction of a$_{1g}$e$_g^\pi$,  
and the band gap $E_{\rm G}=0.6~{\rm eV}$ in antiferromagnetic phase.~\cite{Ezhov-1999} 
The estimated values for the on-site Coulomb and exchange interactions, $U$ and $J$, 
are $2.8$ and $0.93~{\rm eV}$, respectively. 
In the paramagnetic case, the electronic structures by LDA+U calculation is 
quite different from that in the antiferromagnetic case, e.g., 
metallic and large bonding-antibonding 
splitting of a$_{1g}$ orbital which causes 
the majority configuration to be a$_{1g}$e$_g^\pi$.~\cite{Ezhov-1999} 
LDA+DMFT(dynamical mean field theory) calculation 
in paramagnetic phase using the Wannier-type orbitals 
based on NMTO (N-th order muffin-tin orbital method) formalism~\cite{NMTO-2000} 
gives very comprehensive physical picture of V$_2$O$_3$ 
with $U=4.2$~eV, $J=0.7$~eV and the $U^\prime=U-2J =2.8$~eV, 
which presents a narrow resonance peak of a width of 0.5~eV and 
a broad lower Hubbard band.~\cite{V2O3-DMFT} 
A crystal field splitting between a$_{1g}$ and e$_{g}^\pi$ 
is enhanced by the Coulomb interaction, 
then  a$_{1g}$ bands locates  in the large  Hubbard gap  of e$_g^\pi$ orbitals 
and the insulating gap is between the lower e$_g^\pi$ Hubbard band and a$_{1g}$ band. 
Therefore, the insulating gap in the paramagnetic phase 
is the crystal field gap but presumably not the Hubbard gap. 
\begin{table}[t]
\caption{\label{Gap_moment_W_V2O3} The Coulomb and exchange interactions, $U$ (eV) and $J$ (eV), 
 static limit of screened Coulomb interaction $W(0)$ (eV), 
direct band gap $E_{\rm G:d}$ (eV), 
indirect band gap $E_{\rm G:id}$ (eV), 
and the spin magnetic moment $M (\mu_\mathrm B)$ for V$_2$O$_3$. 
The calculated direct and indirect band gaps 
are estimated from the calculated quasiparticle energy. 
}
\begin{ruledtabular}
\begin{tabular}[t]{p{2cm}|cccccc} 
      &  \multicolumn{5}{c}{V$_2$O$_3$} \\
      &$U$ & $J$ &$W(0)$&$E_{\rm G:d}$ &$E_{\rm G:id}$& $M$\\ \hline
LSDA  & -  & -   & -    & (metal) & (metal) &1.40 \\
LSDA+U& 2.6& 0.9 & 2.62 & 0.279   & 0.109   &1.58 \\
GWA   & -  & -   & -    & (metal) & (metal) &1.33 \\
U+GWA & 2.6& 0.9 & 3.21 & 0.943   & 0.835   &1.93 \\
Constrained LSDA\footnotemark & 2.8 & 0.9 & - & - & - & - \\
exp.  &   -&    -&     -& \multicolumn{2}{c}{0.35, 0.66\footnotemark} &1.7\footnotemark  \\ 
\end{tabular}
\end{ruledtabular}
\footnotetext[0]{The value of $W(0)$ depends on the orbital components and 
its averaged one is shown here.}
\footnotetext[1]{I.Solovyev et al. Phys.Rev. B{\bf 53}, 7158 (1996).}
\footnotetext[2]{G.A.Sawatzky and D.Post, Phys. Rev. B{\bf 20}, 1546 (1979); 
G.A.Thomas et al. Phys. Rev. Lett.{\bf 73}, 1529 (1994).}
\footnotetext[3]{Reference \cite{Paolani}}
\label{tab:results_V2O3}
\end{table}%

\begin{figure}[t] 
\begin{center}
\resizebox{0.48\textwidth}{!}{
\includegraphics[height=140mm,clip]{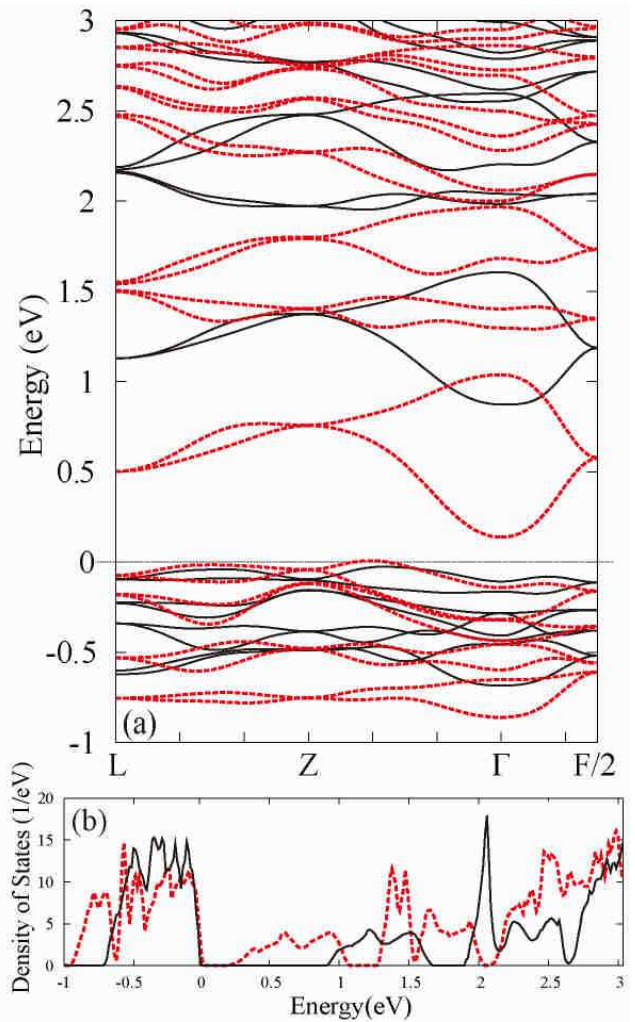}
}
\end{center}
\vspace{-5mm}
\caption{(Color online)
(a) Quasiparticle energy bands and 
(b) the density of states of antiferromagnetic V$_2$O$_3$. 
Solid and broken lines refer to those by U+GWA and LSDA+U 
 with $U=2.6~{\rm eV}$ and $J=0.9~{\rm eV}$. 
The energy zero-th is fixed at the top of the valence band.    
}
\label{fig:band-V2O3}
\end{figure}

\subsection{Results for antiferromagnetic V$_2$O$_3$ by U+GWA}
We use $4 \times 4 \times 4$ ${\bf k}$-point mesh in the Brillouin zone 
of V$_2$O$_3$. 
The set of maximum angular momentum of the LMTO basis in V, O, and two 
different kinds of empty spheres are chosen to be (fdds). 
Those of product basis are (fdds), too, for the calculation of $\Sigma^c$. 
The number of product bases used in $\Sigma^{\rm c}$ 
is reduced from 10,728 to 1,584.
The two crystal structures, monoclinic and corundum structures, 
do not cause any significant difference in the electronic structures  
in LSDA and LSDA+U calculations 
for both paramagnetic and antiferromagnetic phases.  
Then, we use the corundum structure throughout the present work. 
The GWA results of V$_2$O$_3$, starting from LSDA, give rise to 
a metallic antiferromagnetic ground state,  
completely different from the experimental situation.  
Both in LSDA and in LSDA+U, 
the d-bands of majority spin are partially filled 
and those of minority spin are empty. 
Though the screening effects in GWA must be large, 
the polarization function is too large due to small energy denominator, 
similarly to NiO. 
Therefore, we start calculations with LSDA+U to prepare improved wave functions 
and polarization function and adopt U+GWA with the e-only self-consistency. 
The parameters $U$ and $J$ are set as 2.6~eV and 0.9~eV, respectively. 
The value of $U=2.55~{\rm eV}$  may be the lower limit for opening the band gap 
in LSDA+U  
and one would get the metallic antiferromagnetic ground state 
for $U\le2.55~{\rm eV}$.  
The parameters and  calculated results are tabulated 
in Table~\ref{tab:results_V2O3}. 
The Brillouin zones of the paramagnetic and antiferromagnetic phases 
of corundum structure are shown in Fig.~\ref{fig:config-V2O3}(b). 
Figure~\ref{fig:band-V2O3}(a) depicts the quasiparticle band structure 
by U+GWA with $U=2.6~{\rm eV}$ and $J=0.9~{\rm eV}$. 
The ${\bf k}$-point F locates on the surface of the Brillouin zone 
of paramagnetic phase and the mid-point between  F and $\Gamma$ is denoted 
as F/2 locating on that of antiferromagnetic phase. 
The oxygen p-bands with a broad width of 3.7~eV locate at 6~eV below the Fermi energy 
which are not shown here. 
The d-band width becomes narrower due to the formation of a large energy gap. 
The corresponding quasiparticle DOS is shown 
in Fig.~\ref{fig:band-V2O3} (b), compared with DOS by LSDA+U. 
The e$_g^\pi$ bands have a large exchange splitting, 
that of majority spin is almost fully occupied 
and that of minority spin is empty. 
Furthermore, the a$_{1g}$ band of majority spin locates in the gap 
between e$_g^\pi$ majority- and e$_g^\pi$ minority-spin bands. 

\begin{figure}[t] 
\begin{center}
\resizebox{0.49\textwidth}{!}{
\includegraphics[height=7.2cm,clip]{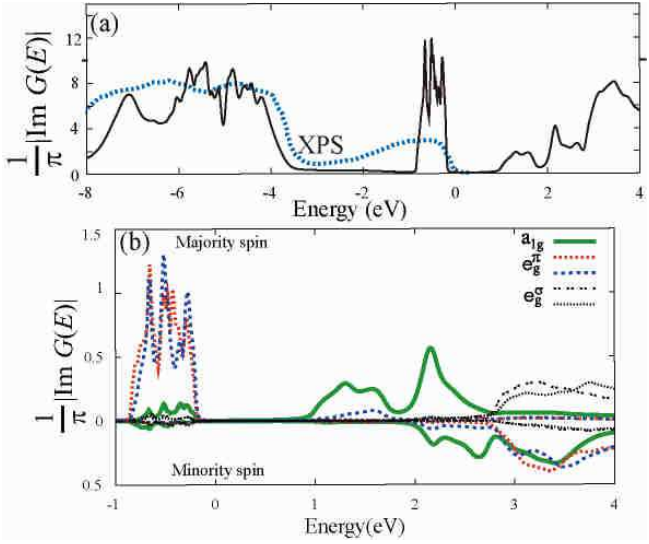}
}
\end{center}
\vspace{-5mm}
\caption{(Color online) 
Imaginary part of the partial Green's functions 
of antiferromagnetic V$_2$O$_3$ by 
U+GWA with $U=2.6~{\rm eV}$ and $J=0.9~{\rm eV}$. 
(a) The total Green's function (solid line) 
and experimental XPS spectrum (dotted line).~\cite{V2O3-exp}
(b) The partial Green's function per atom. 
The energy zero-th is fixed at the top of the valence band. 
a$_{1g}$ and e$_g^\pi$ are originated from d$_{{\rm t}_{1g}}$ orbitals 
in the cubic symmetry  
and e$_g^\sigma$ is originated from d$_{{\rm e}_{g}}$ orbitals. 
}
\label{fig:GreenFn-V2O3}
\end{figure}

The total Green's function $\frac{1}{\pi}|{\rm Im}G|$ is compared with 
the observed XPS spectra in Fig.~\ref{fig:GreenFn-V2O3}(a). 
The position and width of O p-bands are well reproduced in the calculation 
but the width of the occupied V d-bands is too narrow. 
Figure \ref{fig:GreenFn-V2O3}(b) shows the imaginary part of the partial Green's function. 
The valence band just below the Fermi energy 
is mostly of e$_g^\pi$ with small mixing of a$_{1g}$ orbitals and 
the lowest conduction band is of a$_{1g}$ character. 
The ground-state configuration is, then, e$_g^{\pi \sigma}$e$_g^{\pi \sigma}$ of 
majority spins $\sigma$ 
and the e$_g^{\pi -\sigma}$ band of minority spin $-\sigma$ locates 
at higher energies by  about $3.5$~eV. 
The unoccupied a$_{1g}$ band of majority spin is lifted up 
at the midway between majority- and minority-spin e$_g^\pi$ bands.
With an assumption of vanishing hybridization gap, 
the energy difference between main peaks of occupied e$_g^\pi$ and unoccupied a$_{1g}$ 
of parallel spins 
is estimated, replacing $U$ with $U_{\rm U+GWA}=W(0)$ and $J$ by $J_{\rm U+GWA}$, as 
\begin{eqnarray}
&& 
\{E_{\rm C}({\rm e}_g^{\pi\uparrow}{\rm e}_g^{\pi\uparrow}{\rm a}_{1g}^{\uparrow})
-E_{\rm C}({\rm e}_g^{\pi\uparrow}{\rm e}_g^{\pi\uparrow})\} \nonumber \\
&&  \ \ \ \ \ \ \ \ 
-\{E_{\rm C}({\rm e}_g^{\pi\uparrow}{\rm e}_g^{\pi\uparrow})  
-E_{\rm C}({\rm e}_g^{\pi\uparrow})\}   \nonumber \\
&& 
=2(u^\prime_{{\rm t}_{2g}}-j_{{\rm t}_{2g}})-(u^\prime_{{\rm t}_{2g}}-j_{{\rm t}_{2g}})
=W(0)-1.17J_{\rm U+GWA}  \nonumber \\
&& 
\simeq 2.16~{\rm eV}, 
\end{eqnarray}
by using the values used in the U+GWA output, $U_{\rm U+GWA}=W(0)=3.21~{\rm eV}$ and 
$J_{\rm U+GWA}=0.9~{\rm eV}$.   
This result of the energy difference is in good agreement with the calculated one of about 1.5$\sim$2~eV. 
We should note that the gap between majority-spin e$_g^\pi$ and minority-spin e$_g^\pi$ band 
is not the Hund splitting, but is given by the difference in the 
Coulomb interaction (the Coulomb gap);  
\begin{eqnarray}
&& \{E_{\rm C}({\rm e}_g^{\pi\uparrow}{\rm e}_g^{\pi\uparrow}{\rm e}_g^{\pi\downarrow}) 
 -E_{\rm C}({\rm e}_g^{\pi\uparrow}{\rm e}_g^{\pi\uparrow})\} 
\nonumber \\
&&  \ \ \ \ \ \ \ \ 
-\{E_{\rm C}({\rm e}_g^{\pi\uparrow}{\rm e}_g^{\pi\uparrow})
-E_{\rm C}({\rm e}_g^{\pi\uparrow})\} \nonumber \\
&&=(u_{{\rm t}_{2g}}+u_{{\rm t}_{2g}}^\prime)-(u_{{\rm t}_{2g}}^\prime-j_{{\rm t}_{2g}})
 =W(0)+1.91J_{\rm U+GWA}  \nonumber \\
&&
\simeq 4.93~{\rm eV} , 
\end{eqnarray}
which can nicely explain the spectrum in Fig.~\ref{fig:GreenFn-V2O3}.
The position of the minority spin a$_{1g}$ band locates below that of 
the minority spin e$_g^\pi$ band which is estimated as 
\begin{eqnarray}
&& \{E_{\rm C}({\rm e}_g^{\pi\uparrow}{\rm e}_g^{\pi\uparrow}{\rm a}_{1g}^{\pi\downarrow}) 
 -E_{\rm C}({\rm e}_g^{\pi\uparrow}{\rm e}_g^{\pi\uparrow})\} 
\nonumber \\
&&  \ \ \ \ \ \ \ \ 
-\{E_{\rm C}({\rm e}_g^{\pi\uparrow}{\rm e}_g^{\pi\uparrow})
-E_{\rm C}({\rm e}_g^{\pi\uparrow})\} \nonumber \\
&&=(u_{{\rm t}_{2g}}^\prime+u_{{\rm t}_{2g}}^\prime)-(u_{{\rm t}_{2g}}^\prime-j_{{\rm t}_{2g}})
 =W(0)+0.37J_{\rm U+GWA}  \nonumber \\
&&
\simeq 3.54~{\rm eV} . 
\end{eqnarray}
%

The observed V 3d spectrum in paramagnetic metallic phase 
shows a two-peak structure.~\cite{V2O3-exp}
The prominent peak at $E_{\rm F}$ 
of the width of 0.5~eV in metallic phase 
disappears in insulator phase, corresponding to opening the gap, and 
the broader peak with a width of 2~eV does not change much   
among antiferromagnetic and paramagnetic insulator phases.  
The observed V d-bands in insulator phases are the lower Hubbard band and 
may correspond to the occupied U+GWA bands. 
But the present calculated one has a significantly narrower width of about 0.7~eV.  
In fact, the t$_{1g}$ single-electron level locates at 
$u^\prime_{{\rm t}_{2g}}-j_{{\rm t}_{2g}}=W(0)-1.17J_{\rm U+GWA} \simeq 2.16~{\rm eV}$ 
below the lowest $t_{2g}^2$ level 
and would appear between the calculated e$_{g}^\pi$e$_{g}^\pi$ and O 2p-bands. 
Therefore, we believe that the observed broad occupied V d-bands 
originate from the mixture of e$_{g}^\pi$e$_{g}^\pi$ and e$_g^\pi$ states.  
The screened Coulomb interaction can be calculated in GWA and U+GWA. 
Exchange interactions in V$_2$O$_3$ can be calculated also in the framework 
of LSDA+U and $J_\alpha=31~{\rm meV}$ (ferromagnetic), 
$J_\beta=-29~{\rm meV}$ (antiferromagnetic) 
and $J_\gamma=43~{\rm meV}$ (ferromagnetic) 
for the first, second and third nearest neighbor pairs, 
which are consistent with observed spin alignment.
The spin magnetic moment is evaluated to be 1.9~$\mu_{\rm B}$. 
The vanadium trivalent ion is a system with electrons less than half and, therefore, 
the spin-orbit interaction is negative. 
The resultant orbital contribution to the magnetic moment should be negative. 
This is also consistent with experimental observations of the total magnetic moment. 

\subsection{Discussion on paramagnetic V$_2$O$_3$}

In the paramagnetic case, the electronic structure calculated by LDA+U method is 
quite different from that in antiferromagnetic case, e.g., 
metallic and large bonding-antibonding splitting of a$_{1g}$ orbital which causes 
the majority configuration to be a$_{1g}$e$_g^\pi$. 
With a larger value of $U$, the a$_{1g}$ band can move to higher energy region 
but, even if $U=4.0$~eV, the a$_{1g}$ band overlaps still with occupied e$_g^\pi$ band.

The total electron-electron interaction energy can be evaluated  as 
$\frac{1}{3}\{2(u^\prime_{{\rm t}_{2g}}-j_{{\rm t}_{2g}}) +2u^\prime_{{\rm t}_{2g}}
+u_{{\rm t}_{2g}} \}
=1.67W(0)-0.67J_{\rm U+GWA}$
in the case of paramagnetic metallic configuration $({\rm e}_g^{\pi \uparrow})^{1/3}({\rm e}_g^{\pi \uparrow})^{1/3}
({\rm e}_g^{\pi \downarrow})^{1/3}({\rm e}_g^{\pi \downarrow})^{1/3}
(a_{1g}^\uparrow)^{1/3}(a_{1g}^\downarrow)^{1/3}$ 
where we assume that two-electrons occupy the non-degenerate a$_{1g}$ and 
doubly degenerate e$_g^\pi$ bands with spin degeneracy  without the Hubbard gap.   
Similarly we can get  the energy 
$\frac{1}{2}\{(u^\prime_{{\rm t}_{2g}}-j_{{\rm t}_{2g}}) +u^\prime_{{\rm t}_{2g}}
+u_{{\rm t}_{2g}} \}=1.50W(0)-0.21J_{\rm U+GWA}$
in the case of paramagnetic metallic configuration 
$({\rm e}_g^{\pi \uparrow})^{1/2}({\rm e}_g^{\pi \uparrow})^{1/2}
({\rm e}_g^{\pi \downarrow})^{1/2}({\rm e}_g^{\pi \downarrow})^{1/2}$, 
assuming that the a$_{1g}$ band is lifted up and is left empty. 
On the contrary, if we have the Hubbard gap of e$_g^\pi$ bands and 
a$_{1g}$ band is empty, 
then we have a paramagnetic insulator configuration 
e$_g^{\pi \uparrow}$e$_g^{\pi \downarrow}$ 
and the total electron-electron interaction energy can be estimated as 
$u^\prime_{{\rm t}_{2g}}
=W(0)-0.40J_{\rm U+GWA}$. 
We can compare the energies of two configurations where two electrons 
occupy e$_g^\pi$ orbitals, 
the energy difference is 
$(1.5W(0)-0.21J_{\rm U+GWA})-(W(0)-0.4J_{\rm U+GWA})=0.5W(0)+0.19J_{\rm U+GWA}$.
Therefore, the latter configuration 
${\rm e}_g^{\pi \uparrow}{\rm e}_g^{\pi \downarrow}$ should be realized. 
However, the GWA and U+GWA could not create the Hubbard gap 
in paramagnetic phase 
and does not give a realistic feature of electronic structure 
in paramagnetic phase, both insulating and metallic phases. 
This is a possible explanation consistent with LDA+DMFT.~\cite{V2O3-DMFT} 
In the antiferromagnetic case, the situation changes very easily 
because the spin polarization can open the Coulomb gap as we have seen. 
Then the crystal field splitting makes the a$_{1g}$ band appears 
in the insulating gap. 

\section{Summary}\label{Last}
In summary, 
we proposed GWA method starting from the LSDA+U calculation, named U+GWA, 
with energy-only self-consistent calculation. 
The on-site Coulomb interaction parameter is determined 
so that the off-diagonal elements of the self-energy become small 
and we start GWA with more localized wave functions or a wider band gap. 
We then apply U+GWA to antiferromagnetic NiO and V$_2$O$_3$, where 
the LSDA wave functions may be more extended. 
The antiferromagnetic MnO may be a system to which GWA can be applied.  
We have given a general criterion for choosing the on-site Coulomb interaction $U$ 
and the principles whether we should start with LSDA or LSDA+U. 
The band gap, $W(0)$ and spectra for MnO and NiO 
can be evaluated with an excellent agreement 
with the observed results. 
On the contrary, the spectra of V$_2$O$_3$ is much narrower in U+GWA but 
the observed V d-bands may be a mixture of e$_g^\pi$e$_g^\pi$ 
and single-electron e$_g^\pi$ level. 
The method of unique choice of $U$ values has been analyzed in details 
The $U$ values cannot be determined within the U+GWA method but 
with the help of analysis of the XPS and BIS spectra, one can choose a reasonable 
value of $U$ and consistent physical properties can be determined which 
are in excellent agreement with experimental values. 
Computation was partially carried out by use of 
the facilities at the Supercomputer Center, Institute for Solid State Physics, 
University of Tokyo, and the Institute of Molecular Science at Okazaki. 
This work was partially supported by a Grant-in-Aid for Scientific Research in Priority
Areas ``Development of New Quantum Simulators and Quantum Design'' (Grant No. 170640004) of
the Ministry of Education, Culture, Sports, Science, and Technology of Japan.



\begin{thebibliography}{1}

\bibitem{Arai-Fujiwara}
M. Arai and T. Fujiwara, Phys. Rev. B {\bf 51}, 1477 (1995). 

\bibitem{Hedin1965}
L. Hedin, Phys. Rev. {\bf 139}, A796 (1965).

\bibitem{e-only-SC} M. P. Surh, S. G. Louie, and M. L. Cohen, Phys. Rev. B {\bf 43}, 9126 (1991). 

\bibitem{Barth-Holm1996}
 U. von Barth and  B. Holm, Phys. Rev. B {\bf 54}, 8411 (1996); 
 Phys. Rev. B {\bf 55}, 10120 (1997).

\bibitem{Holm-Barth1998}
 B. Holm and U. von Barth, Phys. Rev. B {\bf 57}, 2108 (1998).

\bibitem{Faleev} S. V. Faleev, M. van Schilfgaarde, and T. Kotani, Phys. Rev. Lett. {\bf 93}, 
 126406 (2004).

\bibitem{LSDA+U_3rdGen}
 A. I. Liechtenstein, V. I. Anisimov, and J. Zaanen, Phys. Rev. B{\bf 52}, R5467 (1995).

\bibitem{LSDA+U-LSDA}
 V. I. Anisimov, F. Aryasetiawan, and A. I. Lichtenstein, J. Phys. Condens. Matter {\bf 9}, 767 (1997). 

\bibitem{Miyake2006} Similar method was tried in a paper; 
T. Miyake, P. Zhang, M. L. Cohen, and S. G. Louie, Phys. Rev. B {\bf 74}, 245213 (2006). 
The aim of that paper is just removing the self-interaction of Zn d-orbital 
by using the LDA+U method. 
However, the resultant energy level is very similar to the LDA-starting GWA. 
This may be simply because that the U+GWA does hardly depend on starting Hamiltonian, 
either LDA or LDA+U, 
if the wave functions are enough localized. Certainly this is the case 
for the semi-core d-orbitals of Zn. 
Moreover, GWA is free from the self-interaction even if it starts from the LDA calculation. 
In the present case of NiO or V$_2$O$_3$, 
the aim of starting from LSDA+U calculation is 
to localize d-orbital wave functions but not to remove the self-interaction. 

\bibitem{LMTO}
O.K.~Andersen, Phys. Rev. B{\bf 12}, 3060 (1975);\\
O.K.~Andersen and O.~Jepsen, Phys Rev. Lett. {\bf 53} 2571 (1984); \\
O.K.~Andersen, O.~Jepsen, and D.~Gl\"otzel, in {\it Canonical Description of the Band Structures of Metals}, 
Proceedings of the International School of Physics, ``Enrico Fermi'', Course LXXXIX, Varenna, 1985, 
edited by F.~Bassani, F.~Fumi, and M.P.~Tosi (North-Holland, Amsterdam, 1985), p. 59.


\bibitem{Sakuma2008} 
R. Sakuma, T. Miyake, and F. Aryasetiawan, Phys. Rev. B {\bf 78}, 075106 (2008); 
Usually, one do not use the off-diagonal elements of the self-energy correction 
$\Delta\Sigma_{{\bf k}nn^\prime}$ in the calculation but does actually 
in GWA calculation in VO$_2$.

\bibitem{J-value}
The screened exchange interaction has a small energy dependence but not very much 
different in the series of 3d transition-metal elements. 
Therefore, we fix its value in the present work. 

\bibitem{Ferdi-1}
F. Aryasetiawan and O. Gunnarsson, Phys. Rev. B {\bf 49}, 16214 (1994).

\bibitem{AF-2}
A. Yamasaki and T. Fujiwara, Phys. Rev. B {\bf 66}, 245108 (2002).

\bibitem{MnO-exp_XPS_BIS}
J. van Elp, R. H. Potze, H. Eskes, R. Berger, and G. A. Sawatzky, 
Phys. Rev. B {\bf 44}, 1530 (1991).

\bibitem{Aryasetiawan-NiO-1995}
 F. Aryasetiawan and O. Gunnarsson, Phys. Rev. Lett. {\bf 74}, 3221 (1995).

\bibitem{LDA+U-NiO-1}
 V. I. Anisimov, J. Zaanen, and O. K. Andersen, Phys. Rev. B {\bf 44}, 943 (1991). 

\bibitem{LDA+U-NiO-2}
 V. I. Anisimov, I. V. Solovyev, M. A. Korotin, M. T. Czyzyk and G. A. Sawatzky, Phys. Rev. B {\bf 48}, 16929 (1993). 

\bibitem{Shen}
 Z.-X. Shen, R. S. List, D. S. Dessau, B. O. Wells, O. Jepsen, A. J. Arko, R. Barttlet, 
C. K. Shin, F. Parmigiani, J. C. Huang, and P. A. Lindberg, Phys. Rev. B {\bf 44}, 3604 (1991).

\bibitem{typeNiO} 
 The occupied bands in NiO by the LSDA is mainly Ni d states. 

\bibitem{Paolani}
 L. Paolasini, C. Vettier, F. de Bergevin, F. Yakhou, D. Mannix, A. Stunault, W. Neubeck, 
M. Altarelli, M. Fabrizio, 
P. A. Metcalf, and J. M. Honig, Phys. Rev. Lett. {\bf 82}, 4719 (1999). 
 
\bibitem{Mott1974}
 N. F. Mott, {\it Metal-Insulator Transitions} (Taylor and Francis, London 1974).  

\bibitem{Mattheiss-1994}
 L. F. Mattheiss, J. Phys.:Condens. Matter {\bf 6}, 6477 (1994).

\bibitem{Castellani}
 C. Castellani, C. R. Natoli, and J. Ranninger, Phys. Rev. B {\bf 18}, 4945, 4967, 5001 (1978). 

\bibitem{Ezhov-1999}
 S. Yu. Ezhov, V. I. Anisimov, D. I. Khomskii, and G. A. Sawatzky, Phys. Rev. Lett. {\bf 83}, 4136 (1999). 


\bibitem{NMTO-2000}
O. K. Andersen, and T. Saha-Dasgupta, Phys. Rev. B {\bf 62}, R16129 (2000).

\bibitem{V2O3-DMFT}
A. I. Poteryaev, J. M. Tomczak, S. Biermann, A. Georges, A. I. Lichtenstein, A. N. Rubtsov, 
T. Saha-Dasgupta and O. K. Andersen, Phys. Rev. B{\bf 76}, 085127 (2007).

\bibitem{V2O3-exp} S.-K. Mo, H.-D. Kim, J. D. Denlinger, J. W. Allen, 
J.-H. Park, A. Sekiyama, A. Yamasaki, S. Suga, Y. Saitoh, T. Muro and P. Metcalf, 
Phys. Rev. B{\bf 74}, 165101 (2006). 




\end{thebibliography}
\end{document}